%% file: main-arxiv.tex
\documentclass[sigconf,authorversion,nonacm]{acmart}

\usepackage{caption}
\usepackage{subcaption}
\usepackage{listings}

\lstdefinestyle{mystyle}{
	basicstyle=\scriptsize, 
	breakatwhitespace=false,         
	captionpos=b,                    
	numbers=none, 
	showspaces=false,                
	showstringspaces=false,
	showtabs=false,                  
	tabsize=2,
	frame = single, 
}

\usepackage{enumitem}
\usepackage[nolist]{acronym}
\input{acronym.tex}

\usepackage{cleveref}
\crefname{algocf}{Algorithm}{Algorithms}
\Crefname{algocf}{Algorithm}{Algorithms}

\crefname{section}{Sec.}{Secs.}
\Crefname{section}{Sec.}{Secs.}

\Crefname{equation}{Eq.}{Eqs.}

\crefname{figure}{Figure}{Figures}
\Crefname{figure}{Figure}{Figures}

\AtBeginDocument{%
  \providecommand\BibTeX{{%
    \normalfont B\kern-0.5em{\scshape i\kern-0.25em b}\kern-0.8em\TeX}}}

\copyrightyear{2024}
\acmYear{2024}
\setcopyright{rightsretained}
\acmConference[WiSec '24]{Proceedings of the 17th ACM Conference on Security and Privacy in Wireless and Mobile Networks}{May 27--30, 2024}{Seoul, Republic of Korea}
\acmBooktitle{Proceedings of the 17th ACM Conference on Security and Privacy in Wireless and Mobile Networks (WiSec '24), May 27--30, 2024, Seoul, Republic of Korea}\acmDOI{10.1145/3643833.3656122}
\acmISBN{979-8-4007-0582-3/24/05}

\makeatletter
\gdef\@copyrightpermission{
  \begin{minipage}{0.3\columnwidth}
   \href{https://creativecommons.org/licenses/by-nc-sa/4.0/}{\includegraphics[width=0.90\textwidth]{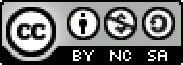}}
  \end{minipage}\hfill
  \begin{minipage}{0.7\columnwidth}
   \href{https://creativecommons.org/licenses/by-nc-sa/4.0/}{This work is licensed under a Creative Commons Attribution-NonCommercial-ShareAlike International 4.0 License.}
  \end{minipage}
  \vspace{5pt}
}
\makeatother

\begin{document}

\title{Over-the-Air Runtime Wi-Fi MAC Address Re-randomization}

\author{Hongyu Jin}
\orcid{0000-0003-2022-3976}
\affiliation{%
	\institution{Networked Systems Security Group\\KTH Royal Institute of Technology}
	\streetaddress{Kistag{\aa}ngen 16}
	\city{Stockholm}
	\postcode{16440}
	\country{Sweden}}
\email{hongyuj@kth.se}

\author{Panos Papadimitratos}
\orcid{0000-0002-3267-5374}
\affiliation{%
	\institution{Networked Systems Security Group\\KTH Royal Institute of Technology}
	\streetaddress{Kistag{\aa}ngen 16}
	\city{Stockholm}
	\postcode{16440}
	\country{Sweden}}
\email{papadim@kth.se}

\renewcommand{\shortauthors}{Hongyu Jin \& Panos Papadimitratos}

\begin{abstract}
	\ac{MAC} address randomization is a key component for privacy protection in Wi-Fi networks.
	Current proposals periodically change the mobile device MAC addresses when it disconnects from the \ac{AP}.
	This way frames cannot be linked across changes, but the mobile device presence is exposed as long as it remains connected: all its communication is trivially linkable by observing the randomized yet same MAC address throughout the connection.
	Our runtime MAC re-randomization scheme addresses this issue, reducing or eliminating Wi-Fi frames linkability without awaiting for or requiring a disconnection.
	Our  MAC re-randomization is practically `over-the-air': MAC addresses are re-randomized just before transmission, while the protocol stacks (at the mobile and the AP) maintain locally the original connection MAC addresses - making our MAC layer scheme transparent to upper layers.
	With an implementation and a set of small-scale experiments with off-the-shelf devices, we show the feasibility of our scheme and the potential towards future deployment.
\end{abstract}

\maketitle

\acresetall
\input{section/introduction}
\input{section/problem}
\input{section/scheme}
\input{section/analysis}
\input{section/implementation}
\input{section/experiment}
\input{section/conclusions}

\section*{Acknowledgments}

This work was supported by the Swedish Research Council project 2020-04621.

\bibliographystyle{ACM-Reference-Format}
\bibliography{references.bib}

\end{document}

%% file: acronym.tex
\begin{acronym}
	\acro{AAA}{Authentication, Authorization and Accounting}
    \acro{AMA}{Adaptive Message Verification}
	\acro{AP}{Access Point}
	\acro{API}{Application Programming Interface}
	\acro{BP}{Baseline Pseudonym}
	\acro{BS}{Base Station} 
    \acro{BSM}{Basic Safety Message}
    \acro{BYOD}{Bring Your Own Device}
   	\acro{C-V2X}{Cellular-V2X}
	\acro{C2C-CC}{Car2Car Communication Consortium}
	\acro{CA}{Certification Authority}
	\acrodefplural{CA}{Certification Authorities}
	\acro{CN}{Common Name}
	\acro{CAM}{Cooperative Awareness Message}
	\acro{CDF}{Cumulative Distribution Function}
	\acro{CIA}{Confidentiality, Integrity and Availability}
	\acro{CRL}{Certificate Revocation List}
	\acro{CDN}{Content Delivery Network}
	\acro{CSR}{Certificate Signing Requests}
	\acro{CTS}{Clear to Send}
	\acro{DA}{Destination Address}
	\acro{DAA}{Direct Anonymous Attestation}
	\acro{DDoS}{Distributed Denial of Service}
	\acro{DDH}{Decisional Diffie-Helman}
	\acro{DENM}{Decentralized Environmental Notification Message}
	\acro{DHT}{Distributed Hash Table}
	\acro{DoS}{Denial of Service}
	\acro{DPA}{Data Protection Agency}
	\acro{DSRC}{Dedicated Short Range Communication}
	\acro{EC}{Elliptic Curve}
	\acro{ECC}{Elliptic Curve Cryptography}
	\acro{ECDSA}{Elliptic Curve Digital Signature Algorithm}
	\acro{ECU}{Electronic Control Unit}
	\acro{ETSI}{European Telecommunications Standards Institute}
	\acro{EVITA}{E-safety Vehicle Intrusion protected Applications}
	\acro{FCFS}{First-Come First-Served}
	\acro{FCS}{Frame Check Sequence}
	\acro{FOT}{Field Operational Test}
	\acro{FPGA}{Field-Programmable Gate Array}
	\acro{GN}{GeoNetworking}
	\acro{GNSS}{Global Navigation Satellite System}
	\acro{GS}{Group Signatures}
	\acro{GM}{Group Manager}
	\acro{GBA}{Generic Bootstrapping Architecture}
	\acro{GPA}{Global Passive Adversary}
	\acro{GUI}{Graphic User Interface}
	\acro{HMAC}{Hash-based Message Authentication Code}
	\acro{HP}{Hybrid Pseudonym}
	\acro{HSM}{Hardware Security Module}
	\acro{HTTP}{Hypertext Transfer Protocol}
	\acro{IEEE}{Institute of Electrical and Electronics Engineers}
	\acro{IoT}{Internet of Things}
	\acro{ITS}{Intelligent Transport System}
	\acro{IT}{Information Technologies}
	\acro{IMSI}{International Mobile Subscriber Identity}
	\acro{IMEI}{International Mobile Station Equipment Identity}
	\acro{IdP}{Identity Provider}
	\acro{ISP}{Internet Service Provider}
	\acro{IV}{Initialization Vector}
	\acro{KRL}{Key Chain Revocation List}
	\acro{LCFS}{Last-Come-First-Served}
	\acro{LEA}{Law Enforcement Agency}
	\acro{LTC}{Long-Term Certificate}
	\acro{LTCA}{Long-Term \acl{CA}}
	\acrodefplural{LTCA}{Long-Term \aclp{CA}}
	\acro{LDAP}{Lightweight Directory Access Protocol}
	\acro{LTE}{Long Term Evolution}
	\acro{LBS}{Location-based Service}
	\acro{MAC}{Medium Access Control}
	\acro{MCA}{Message \ac{CA}}
	\acro{MEA}{Misbehavior Evaluation Authority}
	\acro{NIC}{Network Interface Card}
	\acro{NTP}{Network Time Protocol}
	\acro{OBU}{On-Board Unit}
	\acro{OCSP}{Online Certificate Status Protocol}
	\acro{OSN}{Online Social Network}
	\acro{PC}{Pseudonymous Certificate}
	\acro{PCA}{Pseudonymous \acl{CA}}
	\acrodefplural{PCA}{Pseudonymous \aclp{CA}}
	\acro{PDP}{Policy Decision Point}
	\acro{PEP}{Policy Enforcement Point}
 	\acro{PET}{Privacy-enhancing Technology}
  	\acrodefplural{PET}{Privacy-enhancing Technologies}
	\acro{PIR}{Private Information Retrieval}
	
	\acro{PKI}{Public-Key Infrastructure}
	\acro{PN}{Packet Number}
	\acro{POI}{Point of Interest}
	\acro{PRECIOSA}{Privacy Enabled Capability in Co-operative Systems and Safety Applications}
	\acro{PRESERVE}{Preparing Secure Vehicle-to-X Communication Systems}
	\acro{PRL}{Pseudonymous Certificate Revocation List}
	\acro{PTK}{Pairwise Transient Key}
	\acro{P2P}{peer-to-peer}
	\acro{RA}{Receiver Address}
	\acro{REST}{Representational State Transfer}
	\acro{RBAC}{Role Based Access Control}
	\acro{RCA}{Root \acl{CA}}
	\acro{RSU}{Roadside Unit}
	\acro{RTS}{Request to Send}
	\acro{SA}{Source Address}
	\acro{SAE}{Simultaneous Authentication of Equals}
	\acro{SAML}{Security Assertion Markup Language}
	\acro{SCORE@F}{Système COopératif Routier Expérimental Français}
	\acro{SeVeCom}{Secure Vehicle Communication}
	\acro{SIS}{Security Infrastructure Server}
	\acro{SN}{Sequence Number}
	\acro{SP}{Service Provider}
	\acro{SSO}{Single-Sign-On}
	\acro{SoA}{Service-oriented-Approach}
	\acro{SOAP}{Simple Object Access Protocol}
	\acro{SAS}{Sample Aggregation Service}
	\acro{TA}{Transmitter Address}
	\acro{TESLA}{Timed Efficient Stream Loss-Tolerant Authentication}
	\acro{TS}{Task Service}
	\acro{TLS}{Transport Layer Security}
	\acro{TPM}{Trusted Platform Module}
	\acro{TTP}{Trusted Third Party}
	\acro{TVR}{Ticket Validation Repository}
	\acro{URI}{Uniform Resource Identifier}
	\acro{VANET}{Vehicular Ad-hoc Network}
	\acro{V2I}{Vehicle-to-Infrastructure}
	\acro{V2V}{Vehicle-to-Vehicle}
	\acro{V2X}{Vehicle-to-Everything}
	\acro{VC}{Vehicular Communication}
	\acro{VM}{Virtual Machine}
	\acro{VSN}{Vehicular Social Network}
	\acro{VSS}{\ac{VC} Security Subsystem}
	\acro{WAVE}{Wireless Access in Vehicular Environments}
	\acro{WSDL}{Web Services Discovery Language}
	\acro{W3C}{World Wide Web Consortium}
	\acro{VANET}{Vehicular Ad-hoc Network}
	\acro{VPKI}{Vehicular Public-Key Infrastructure}
	\acro{WS}{Web Service}
	\acro{WoT}{Web of Trust}
	\acro{WSACA}{\ac{WAVE} Service Advertisement \ac{CA}}
	\acro{XML}{Extensible Markup Language}
	\acro{XACML}{eXtensible Access Control Markup Language}
	\acro{3G}{3rd Generation}
\end{acronym}

%% file: section/introduction.tex
\section{Introduction}

Wi-Fi networks are a corner-stone for Internet access, at homes, offices, and public areas, enabling seamless connectivity for laptops, smartphones, smart wearables, etc., each identified by a \ac{MAC} address, a globally unique identifier assigned to each \ac{NIC}.
\ac{MAC} addresses are hard-coded by the \ac{NIC} manufacturers.
When a device transmits Wi-Fi frames, it specifies its \ac{MAC} address in the \ac{SA}/\ac{TA} field in the frame header~\cite{ieee80211}.
Transmitted (Tx) Wi-Fi frames can be easily monitored by \acp{NIC} that can operate in `monitor/promiscuous' mode.
A monitoring device can track devices in an area by observing received (Rx) Wi-Fi frames with the same SA/TA \ac{MAC} address.
Multiple monitoring devices can track devices (and, consequently, their users) across numerous points-of-interest targeted users (devices) are likely to visit.

\ac{MAC} address randomization~\cite{fenske2021three,ietf-madinas-mac-address-randomization-09} emerged as a countermeasure to such so-called sniffing attacks.
Instead of the hard-coded \ac{MAC} address, the driver/firmware reads software-generated \ac{MAC} addresses, refreshed periodically or based on specific conditions (e.g., per SSID or per connection).
With encryption covering Wi-Fi frame payloads, device transmissions are still easily linkable while the same MAC address is used.
Once the device is assigned a new randomized MAC address, a local eavesdropper cannot link traffic to that by the same device with an earlier randomized MAC address.
Thus, such relatively shorter-term MAC addresses significantly reduce linkability of traffic and thus user presence/activities.

The level of privacy protection (unlinkability) is not high: while connected to an \ac{AP}, given the device uses the same MAC address throughout the connection, inference attacks on user activities are possible.
More frequent MAC randomization is possible by manually disconnecting and re-connecting to \acp{AP}, at the expense of inefficient communication, without fundamentally solving the problem at hand (see \cref{sec:problem}).

This motivates our proposal of over-the-air runtime MAC re-randomization, that is, MAC randomization within and repeatedly throughout the connection/session without affecting the communication quality, i.e., without disconnections and no data rate or reliability reduction.
Our scheme only randomizes the MAC addresses in the headers of Tx frames on the verge of transmission, while the Rx devices recover the original/actual (termed \emph{base}) MAC addresses immediately upon frame reception.
This makes our scheme transparent to upper layer protocols.
At the same time, an eavesdropper only sees the ephemeral re-randomized MAC addresses in the Tx frame headers.
The re-randomization is performed synchronously by all stations connected to an AP, drastically growing the anonymity set to be equivalent to the set of connected devices.

We provide a detailed description of our scheme, including \ac{SN} and nonce (this term is used interchangeably with \ac{PN} and \ac{IV}) reset approaches (\cref{sec:scheme}), and analyze its security and privacy (\cref{sec:analysis}).
We implement our scheme with off-the-shelf Wi-Fi \acp{NIC} at the driver level (\cref{sec:implementation}).
A set of small-scale experimental results shows that our scheme does not deteriorate communication performance (\cref{sec:evaluation}), before our concluding remarks (\cref{sec:conclusion}).

\textbf{Related Works:} To the best of our knowledge, the only existing work that addresses MAC address rotation during Wi-Fi connection is RoMA~\cite{hugon2022roma}.
However, RoMA does not provide seamless MAC address rotation from the perspective of the \ac{AP}, because different IP addresses are used for virtual interfaces; the number of rotating MAC addresses during a connection is limited by the maximum number of supported Wi-Fi NIC virtual interfaces.
Cryptographically generated~\cite{greenstein2008improving} randomized addresses were not Wi-Fi compatible.
The current MAC randomization protocol, work-in-progress with an IETF draft~\cite{ietf-madinas-mac-address-randomization-09}, defines 46-bit randomization (with 2 fixed bits defining unicast and locally administered addresses).
The latest OSes implement their variants of MAC randomization, with major differences reflected on randomized MAC address lifetime (e.g., per SSID or per connection)~\cite{ietf-madinas-mac-address-randomization-09,martin2017study,fenske2021three}.
We refer readers to~\cite{martin2017study,fenske2021three} for a detailed introduction to MAC randomization policies.

%% file: section/problem.tex
\section{Problem Statement}
\label{sec:problem}

\ac{MAC} randomization has been adopted by the latest mobile devices (e.g., Apple and Android based~\cite{applemac,androidmac}) to prevent tracking connections to different APs/SSIDs in public areas.
However, the same randomized MAC address is used throughout a connection.
This way, an eavesdropping/sniffing attacker can easily identify a device (period of) presence: by eavesdropping the appearance and disappearance of a specific mobile MAC frame address in the network.
For example, a person at work connects her/his device to an AP in the office area having the same MAC address throughout the day.
Therefore, randomized \ac{MAC} addresses do not prevent inferring user activity or linking the same randomized \ac{MAC} address over different connections (e.g., over multiple days).

A straightforward improvement is to periodically force disconnection from the AP and re-connection with a fresh randomized \ac{MAC} address; triggered manually by the user or automatically by the system.
However, such an approach would not be very effective because the attacker could easily link the two (old and new) MAC addresses by observing the appearance of a new MAC address in the network immediately after the disappearance of an old MAC address.

Inspired by the mix-zone approach~\cite{beresford2003location,Papadimitratos2019} for identity and location protection, where users change pseudonyms in a mix-zone synchronously, devices connected to an AP can synchronize their disconnection and re-connection with freshly randomized MAC addresses.
Such synchronized actions can be performed periodically to render MAC address linking based on the appearance/disappearance timing impossible; because multiple MAC addresses would disappear and appear at the same time.
However, such forced dis- and re-connections could cause substantial delays at each re-connection.
Moreover, some devices might refuse to disconnect and change \ac{MAC} address at a given point - e.g., due to ongoing communication (e.g., important video calls) of critical nature - at the expense of remaining linkable. 
Having several devices refraining from changing their MAC addresses shrinks the anonymity set for devices that did change their address.

These limitations motivate us to propose the runtime MAC re-randomization scheme described in \Cref{sec:scheme}: an effective scheme allowing fast changes across randomized MAC addresses without degrading communication performance, and preventing frame linking based on continuous \acfp{SN} and WPA2/3 nonces.
Our scheme protects against any eavesdroppers (including devices connected to the same AP) and retains the same level of privacy protection against the connected AP as that of classic MAC randomization~\cite{ietf-madinas-mac-address-randomization-09,martin2017study,fenske2021three}.
Fingerprinting attacks based on traffic analysis~\cite{vanhoef2016mac,matte2016defeating,martin2017study} are out of the scope of this paper and part of future work.

%% file: section/scheme.tex
\section{Our Scheme}
\label{sec:scheme}

We build on top of MAC randomization, to ensure that a roaming user/mobile device (termed \emph{station}) that connects to a new AP selects a new randomly chosen MAC address.
We term this the \emph{base MAC address} for the current connection to the AP.
The base address of every station is known and kept by the AP, clearly, not needing to change/randomize further.
The first novelty of our scheme is to not use the base address but rather assign a new periodically \emph{re-randomized MAC address} for Tx frames.
Re-randomized addresses are computed with a hash function based on a common secret, the WPA2/3 \ac{PTK}, between each station and the AP (\cref{sec:computation}).
Computations are performed on both the AP and each station independently, based on identical inputs, thus resulting in identical new, re-randomized MAC addresses.
For each new re-randomization, \acp{SN} and WPA2/3 nonces are also reset to prevent linking based on these (otherwise continuous) values (\cref{sec:reset}).
In order to prevent nonce reuse, we propose a controlled nonce reset to make sure the 48-bit nonce wrap interval is significantly longer than a \ac{PTK} lifetime.
Our scheme is transparent to the upper layers because it ensures that only the re-randomized MAC addresses are exposed over the air, while the protocol stacks operate with the base MAC address (\cref{sec:spoofing}).

\subsection{MAC Computation}
\label{sec:computation}

Our scheme assumes the base MAC address randomization is handled by the station itself.
This can be easily supported by mobile devices nowadays.
For example, the \textit{macchanger} tool can randomize the MAC address in Linux systems.
Although current default Apple and Android policies don't change MAC addresses for each new connection to the same SSID without user intervention (e.g., by `forgetting' the network or resetting the network settings)~\cite{applemac,androidmac}, it is clear that this can be supported (currently restricted by system policies).

After the \ac{AP} and a station are connected (thus, shared keys computed), the \ac{MAC} re-randomization is performed periodically, every $T$ seconds; $T$ is the lifetime of each re-randomized MAC address.
Transitions to a new MAC address are synchronized across all stations connected to an AP.
The computation of the re-randomized \ac{MAC} address is based on a hash function with three input values, carried out by the AP and each station locally, based on the corresponding pair-wise shared secret (key).

\begin{itemize}[leftmargin=*]
	\item The first input is the base (static) \ac{MAC} address, used to initiate the current connection to the AP. 
	\item The second input is the \ac{PTK} computed during the authentication phase~\cite{wpa3}, only known by the \ac{AP} and the corresponding station, thus the shared secret for each AP-station pair.
	In WPA2, a \ac{PTK} (thus the re-randomized MAC addresses) can be derived by any eavesdropper that monitored a connection from its very beginning (i.e., recorded all handshake frames) and knows the passphrase.
	Thus, WPA2 is not fully secure per se and cannot be used to enhance privacy beyond the standard MAC randomization approach~\cite{ietf-madinas-mac-address-randomization-09,martin2017study,fenske2021three} with our scheme.
	However, the Dragonfly/\ac{SAE} handshake protocol in WPA3 requires active participation to compute the \ac{PTK}, essentially preventing the eavesdropper from computing the \ac{PTK} and consequently the re-randomized MAC addresses of other stations, even if she were present from the beginning of their connection~\cite{wpa3}.
	\item The last input is a unique index of the $T$ interval, i.e., the validity period of the re-randomized MAC address. The index is computed with $\lfloor t_{epoch} / T \rfloor$, where $t_{epoch}$ is the Unix (or epoch) time in seconds (counted from January 1, 1970).
\end{itemize}

The three values are concatenated and hashed to produce a digest whose first six bytes are used as the basis for a random \ac{MAC} address.
In order to derive a valid MAC address, we have to set to the first byte bit-0=0 (unicast) and bit-1=1 (locally administered)~\cite{ietf-madinas-mac-address-randomization-09,ieee80211}.
With this adjustment, the six bytes are ready to be used as the newly re-randomized \ac{MAC} address.
Each station only needs to maintain the currently used re-randomized \ac{MAC} address (for the current $T$ interval) together with the base address, and convert between the base one and the re-randomized one during frame transmission and reception.
The AP needs to (i) compute for each station its re-randomized MAC address using the same inputs and function as described above, and (ii) maintain a table of base and re-randomized MAC addresses for MAC conversion for Wi-Fi frames sent/received to/from the stations.
The AP does not need to randomize (let alone re-randomize) its MAC address because there is no need to protect the AP privacy.

\subsection{Sequence Number and Nonce Reset}
\label{sec:reset}

To prevent linking Wi-Fi frames based on their continuous \acp{SN} and nonces, we have to reset them along with each MAC address re-randomization.
The \ac{SN} is simply reset to 0, a standard operation that usually happens when the 12-bit incremental \ac{SN} reaches its maximum value (i.e., 4095; or 2047 allowed on some devices~\cite{fenske2021three}).
In our scheme, the \ac{SN} reset happens together with the \ac{MAC} re-randomization too.

In WPA2/3, 48-bit nonces (essentially \acp{PN}) are used to ensure the same plaintext does not result in the same ciphertext~\cite{vanhoef2018release}.
Unlike \acp{SN}, resetting the nonce to 0 would create a vulnerability, due to nonce reuse.
A straightforward approach is to set the nonce to a random value with the new MAC address, and increment as usual, from that point until the next MAC re-randomization.
The large enough 48-bit nonce space would naturally render nonce reuse highly unlikely.
This can be augmented by checking whether frames within the upcoming $T$ interval could potentially reuse earlier own nonces (depending on the bit rate and frame size) and randomize again if needed.
This has a privacy implication: frames using previously seen nonces shrink the anonymity set because frames with identical nonces do not originate from the same station (MAC addresses).

\begin{figure}[h]
	\centering
	\includegraphics[width=0.5\columnwidth]{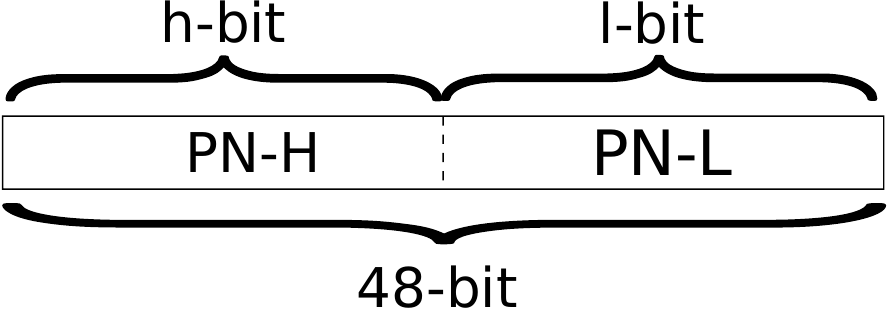}
	\caption{48-bit nonce/PN comprises $h$/$l$-bit PN-H/PN-L.}
	\label{fig:pn}
\end{figure}

To thwart nonce reuse and the potential privacy issue above, we propose a controlled nonce/PN reset approach.
As shown in \cref{fig:pn}, we split the PN field into PN-H and PN-L, $h$-bit and $l$-bit long, respectively, where $h+l=48$.
For the corresponding $T$ interval, we set $\text{PN-H} = \lfloor t_{epoch} / T \rfloor \ mod\ 2^h$, and PN-L is simply reset to 0 with each new MAC re-randomization.
In other words, PN-H is incremented by 1 every $T$, and PN-L is the nonce space used for the frames within the $T$ interval.
$h$ and $l$ values can be chosen based on \Cref{eq:l}, where $L_{frame}$ is the average frame size, and the interval between two PN wraps is $2^{h} * T$.
We show in \cref{sec:analysis} that nonce reuse is essentially impossible with practical frame sizes and bit rates.

\begin{align}
	l = \lceil log_2(\frac{Bitrate * T}{L_{frame}}) \rceil, \ h = 48-l \label{eq:l}
\end{align}

\subsection{Over-the-Air MAC Conversion}
\label{sec:spoofing}

\begin{figure}[h]
	\centering
	\includegraphics[width=\columnwidth]{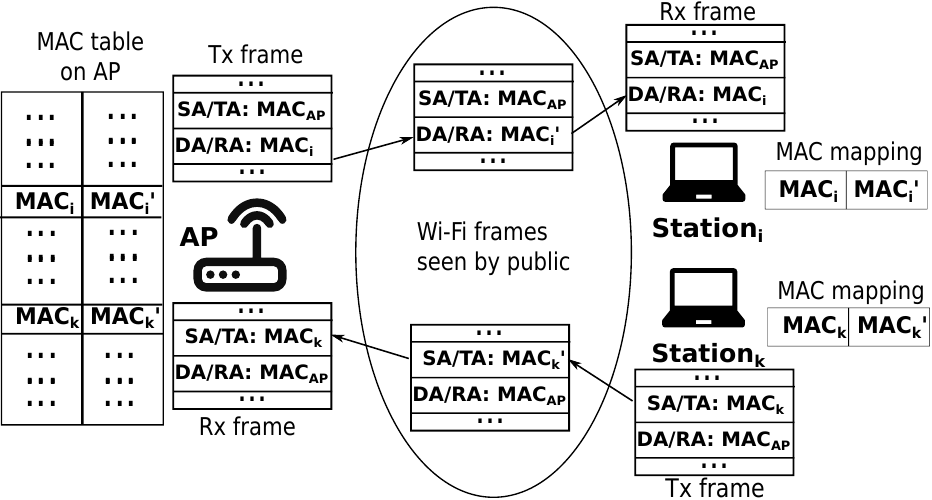}
	\caption{Over-the-air MAC conversion.}
	\label{fig:mac}
\end{figure}

\Cref{fig:mac} illustrates the over-the-air MAC conversion.
The AP maintains a MAC address table, one row per connected mobile station; with the base MAC addresses used for establishing connections in the left column and the currently valid re-randomized MAC addresses in the right column.
Immediately before transmitting a Tx frame, the AP performs a lookup in the MAC address table with the \ac{DA}/\ac{RA} address (a base address).
If there is a match, the \ac{DA}/\ac{RA} address is converted to the corresponding re-randomized MAC address.
For an Rx frame, the AP lookup is carried out with the \ac{SA}/\ac{TA} address at the right column of the MAC table, immediately upon Rx frame reception.
If there is a match, the \ac{SA}/\ac{TA} address is converted to the corresponding base MAC address, transparently to the upper layers, so that the Rx frame payload is properly decapsulated.
Each station only maintains its own re-randomized MAC addresses, apart from the base address.
Similarly, each station converts the \ac{SA}/\ac{TA} address on a Tx frame and the \ac{DA}/\ac{RA} address (that matches own re-randomized addresses) on an Rx frame.
In summary, only the re-randomized MAC addresses are exposed over the air while the AP and station protocol stacks use the station base MAC address.
The MAC address conversion has to be performed before the \ac{FCS} is calculated, usually done at the hardware level; otherwise, the corresponding frames would be deemed corrupted and dropped by receivers.

%% file: section/analysis.tex
\section{Security and Privacy Analysis}
\label{sec:analysis}

We provide a security analysis on nonce reuse and a privacy analysis on the unlinkability of MAC addresses, based on the unencrypted MAC header fields over the synchronized re-randomizations.
Traffic (metadata) analysis or device fingerprinting are out of scope.

\subsection{Preventing Nonce Reuse}

\begin{figure}[t]
	\centering
	\includegraphics[width=0.5\columnwidth]{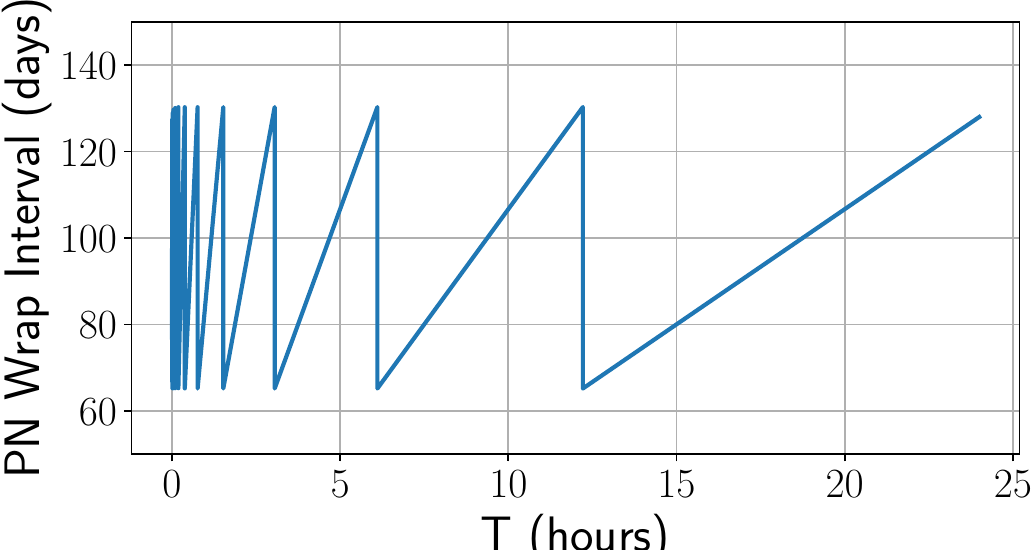}
	\caption{Interval between two PN wraps, as a function of $T$ (1$s$ $\sim$ 24$h$), for 50 $byte$ data frames and 10 $Gbps$ bit rate.}
	\label{fig:wrap}
\end{figure}

Nonce reuse with a same \ac{PTK} is prevented by choosing proper $h$ and $l$ values based on system parameters (\Cref{eq:l}).
We explain, with the help of a demanding example, why nonce reuse is practically impossible with our scheme.
\Cref{fig:wrap} shows the interval between two nonce wraps (i.e., PN-H wraps to zero) considering two extreme values for average data frame size (50 $bytes$, near minimal) and bit rate (10 $Gbps$).
Even so, the wrap happens every 60 days or more, much longer than the expected lifetime of a \ac{PTK}.
With larger frame sizes and lower bit rates, the number of theoretically possible frame transmissions would be lower.
This implies lower $l$ and higher $h$ values, thus even longer wrap intervals than those in \cref{fig:wrap}.

\subsection{MAC Address Unlinkability}

\begin{figure}[t]
	\centering
	\includegraphics[width=0.6\columnwidth]{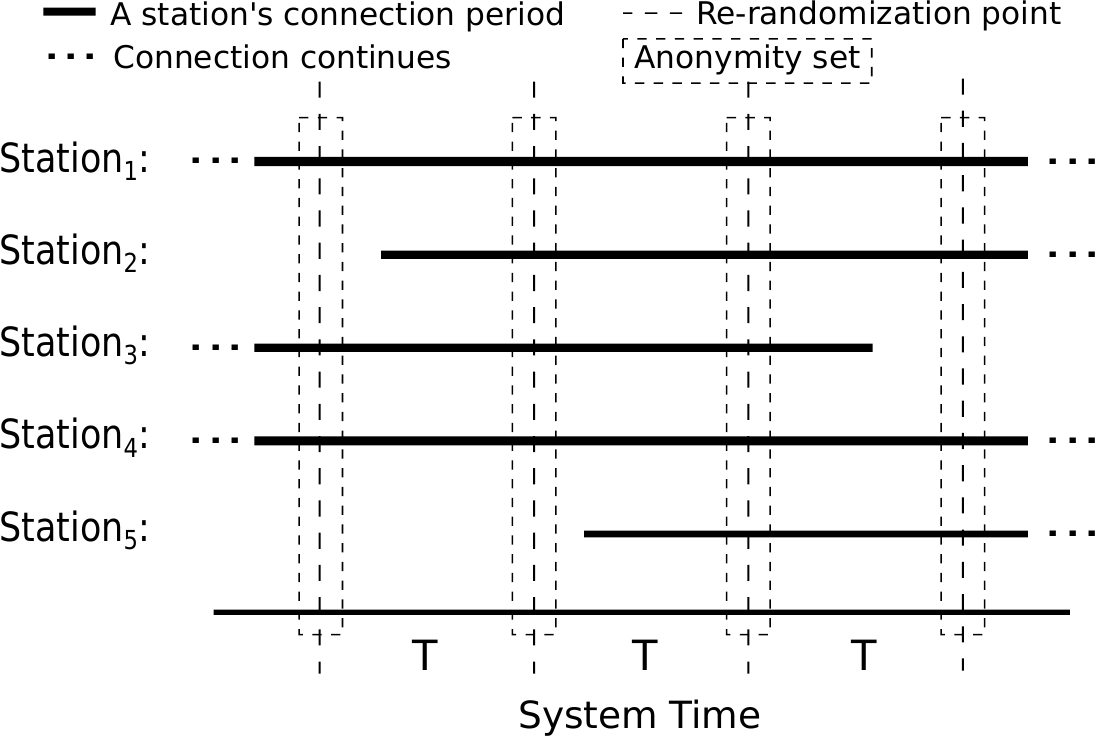}
	\caption{MAC re-randomization example with five stations.}
	\label{fig:rand}
\end{figure}

\Cref{fig:rand} shows an example of MAC re-randomization for five station connections.
As per \cref{sec:problem}, if stations re-randomize/change their MAC address independently, thus in all likelihood at distinct points in time, an eavesdropper could easily link any two consecutive MAC addresses of each station (possibly based on continuing activity across the MAC address change).
However, with our synchronized MAC re-randomization, the anonymity set after each MAC change would be the same as the number of connected stations; thus linking the old and new MAC addresses of the same device would be hard.
The initial connection to the AP always stands out due to the handshake frames, but the subsequent frames will not be linkable based on MAC addresses.
Hiding a disconnection is possible if a station leaves the network without sending disconnection frames and $T$ is relatively short; thus the eavesdropper would not know whether the station left or it was just idle, without any communication for a short period until the next re-randomization point.
This would incur slightly more overhead on the AP, needing to maintain the disconnected stations until their removal due to inactivity timeouts.
Moreover, unified \ac{SN} and nonce resets (i.e., reset to identical values by each station and the AP) make linking based on these (otherwise continuous) values impossible.
Last but not least, we emphasize our scheme does not enhance privacy against connected \acp{AP}: they learn the same information the standard MAC randomization provides.

%% file: section/implementation.tex
\section{Implementation}
\label{sec:implementation}

We implement our MAC re-randomization and run experiments on four identical small form factor computers with Atheros AR5414 \acp{NIC}~\cite{ar5414}, Intel Atom D510 dual-core 1.66GHz CPUs, and 1 GB RAM.
They are all installed with Debian OS 11.8, with Linux kernel version 5.10.0-26.
AR5414 is a legacy \ac{NIC} model that supports only IEEE 802.11 a/b/g, not the latest IEEE 802.11 n/ac/ax (Wi-Fi 4/5/6).
However, we choose this available to us model due to Linux support through the fully open-source driver, ath5k, without any non-free binary firmware.
Therefore, we have full control of the device through its driver.
This makes AR5414 perfect for testing the feasibility of our approach, as a first step towards a larger scale deployment and experiments with the latest devices.
We implement our scheme by modifying the ath5k driver and the mac80211 module provided through Linux backports driver version 5.10.168\footnote{\url{https://cdn.kernel.org/pub/linux/kernel/projects/backports/stable/}}.

\subsection{MAC Re-randomization}

Our scheme requires each station to randomize its base MAC address used for connection through system tools, e.g., $macchanger$, to set a randomized MAC address for the wireless interface in Linux.
For the re-randomized \ac{MAC} computation after the established connection, we use the SHA256 function from the Linux kernel crypto module to compute hash digests, as described in \cref{sec:scheme}, retaining the first six bytes.
Then, we set bit-0=0 and bit-1=1 of the first byte respectively, to ensure the MAC address is valid.
At the \ac{AP}, we use the Linux kernel hash table\footnote{\url{https://lwn.net/Articles/510202/}} to maintain the MAC address table.
For each station, the AP maintains two entries (i.e., key-value pairs) in its hash table.
The keys of the two entries are the base MAC and the re-randomized MAC addresses for each connected station.
The values of the two entries are identical: a $struct$ comprises both the base and the re-randomized MAC addresses.
The two entries are used to look up MAC addresses to convert to for Tx or Rx frames.
We declare a 10-bit hash table in our implementation, storing $2^{10}$ entries for at most 512 stations.
With AR5414, \ac{FCS} can be correctly calculated by the hardware after the Tx frame MAC address conversion by the driver.

\subsection{Resetting Sequence Number and Nonce}

We found that the AR5414 would assign \acp{SN} on the hardware by overwriting the driver-assigned \acp{SN}.
Therefore, we had to disable hardware-assigned \acp{SN} to respect the driver-assigned \acp{SN} (\cref{sec:reset}).
We were able to reset nonces in the driver without any issue.

\subsection{Enabling ACK/RTS/CTS Frames}

\subsubsection{ACK Frames}

Time-sensitive ACK frames acknowledge data frames, triggered upon reception by the hardware, not the driver (the hardware-driver communication would introduce extra delay).
Usually, \acp{NIC} in \emph{station} mode only acknowledge frames that specify \ac{RA}/\ac{DA} as the Rx station local addresses.
In our implementation, given the transmitted frames carry the re-randomized \ac{MAC} addresses that do not match the local addresses, the Rx hardware would not trigger ACK frames, resulting in a series of re-transmissions of non-ACKed data frames.

This can be solved by setting a BSSID mask (originally used to enable multiple virtual interfaces) on AR5414~\cite{ar5414}: after each MAC re-randomization, setting a BSSID mask (computed with inputs the base and the newly re-randomized \ac{MAC} addresses) ensures ACK frames are properly triggered.
However, this opens up a privacy issue.
An attacker could inject frames by specifying the (eavesdropped) base MAC address of a station as the \ac{DA}/\ac{RA}.
These injected frames would trigger ACK frames and expose the presence of the device with the targeted (sought by the adversary) base MAC address.

In order to fundamentally solve the problem, we directly update the main MAC address register~\cite{ar5414} on the hardware with the re-randomized MAC address.
This allows ACK frames to be triggered only for the currently used re-randomized MAC address, not for the base MAC address.
This is the expected correct behavior of our scheme: stations should not respond to any frame destined for their base MAC addresses.
Otherwise, the presence of the device with the specific base MAC address would be exposed.
 
\subsubsection{RTS/CTS Frames}

Similar to ACK frames, time-sensitive \ac{RTS}/\ac{CTS} frames\footnote{They are intended to make the shared medium access more efficient, solving the hidden and exposed terminal problems.} are also handled by the hardware~\cite{ar5414}.
A connected station sends an \ac{RTS}, awaits for a \ac{CTS} sent in response by the AP and only then transmits data.
Most APs would disable RTS/CTS frames due to performance uncertainty; e.g., OpenWrt disables RTS/CTS (confirmed by installing the OpenWrt firmware on two different ASUS router models), and dd-wrt sets the default RTS threshold to 2347 $bytes$\footnote{\url{https://wiki.dd-wrt.com/wiki/index.php/Advanced_wireless_settings}} (i.e., effectively disabled, given the maximum Wi-Fi frame size being 2346 $bytes$).
However, in order to demonstrate the usability of RTS/CTS frames in our implementation, we do enable RTS/CTS frames by setting a moderate threshold on the stations.
We see RTS/CTS frames properly triggered with the re-randomized \ac{MAC} addresses written to the hardware register (see \Cref{sec:evaluation}).

%% file: section/experiment.tex
\section{Experimental Evaluation}
\label{sec:evaluation}

We find that our scheme and the modified driver do not deteriorate the communication performance in the experimental setup.
Due to the legacy hardware and the limited number of devices available, we do not present an extensive evaluation.
Our goal is to show the feasibility of our scheme by comparing with the vanilla driver, serving as a stepping stone to a larger scale experiment with the latest devices in the future.

\subsection{Experimental Setup}

One device acts as the \ac{AP} and the other three act as stations, deployed in a small office.
$Station_1$ is in line-of-sight, and $Station_2$ and $Station_3$ are in non-line-of-sight of the AP.
We use $hostapd$ to set up the \ac{AP} and $wpa\_supplicant$ for the stations.
The AP is configured with WPA3-Personal.
The stations are assigned static IPv4 addresses, and access the Internet wirelessly, routed through the AP.
We set $T=30s$, and both $h$ and $l$ to 24, thus 24-bit PN-H/L.
The values are sufficient to prevent nonce reuse for the bit rate supported by the devices.
We provide the full $hostapd$ and $wpa\_suppplicant$ configurations below.

\begin{minipage}[b]{.45\columnwidth}
	\begin{lstlisting}[caption=$hostapd$,mathescape=true]
		interface=<INTERFACE>
		driver=nl80211
		ssid=<SSID>
		hw_mode=g
		channel=9
		$\textbf{\# The next two lines enable WPA3}$
		wpa=2
		wpa_key_mgmt=SAE
		rsn_pairwise=CCMP
		wpa_passphrase=<PASSWORD>
		ieee80211w=2
	\end{lstlisting}
\end{minipage}\hfill
\begin{minipage}[b]{.45\columnwidth}
	\begin{lstlisting}[caption=$wpa\_supplicant$]
		network={
			ssid=<SSID>
			psk=<PASSWORD>
			key_mgmt=SAE
			ieee80211w=2
		}
	\end{lstlisting}
\end{minipage}\hfill

\subsection{Wi-Fi Frame Monitoring}

We also place another device in the same space, with its wireless interface set to \emph{monitor} mode on the same channel as the AP and three stations.
Real-time monitoring data through Wireshark\footnote{\url{https://www.wireshark.org/}} confirms that the MAC addresses are refreshed every $T=30s$.
Once a new $T$ interval is reached, older MAC addresses no longer show up in the captured frames, while the ongoing communication is undisrupted.
We also observe that ACK/RTS/CTS frames, handled by the hardware, are properly triggered thanks to the hardware overwritten MAC address registers.

\subsection{Performance Evaluation}

In order to compare the performance of the vanilla driver and our implementation, we conduct several experiments, copying files between a local server and the stations, and downloading files from the Debian website.
We repeatedly run $scp$ copy or $wget$ download commands (i.e., a new command is executed after the previous one concludes), to make sure all three stations communicate with the AP simultaneously.
We show the average speed for five operations (commands run) for each station.

\begin{figure}[t]
	\centering
	\begin{subfigure}[b]{0.45\columnwidth}
		\includegraphics[width=\columnwidth]{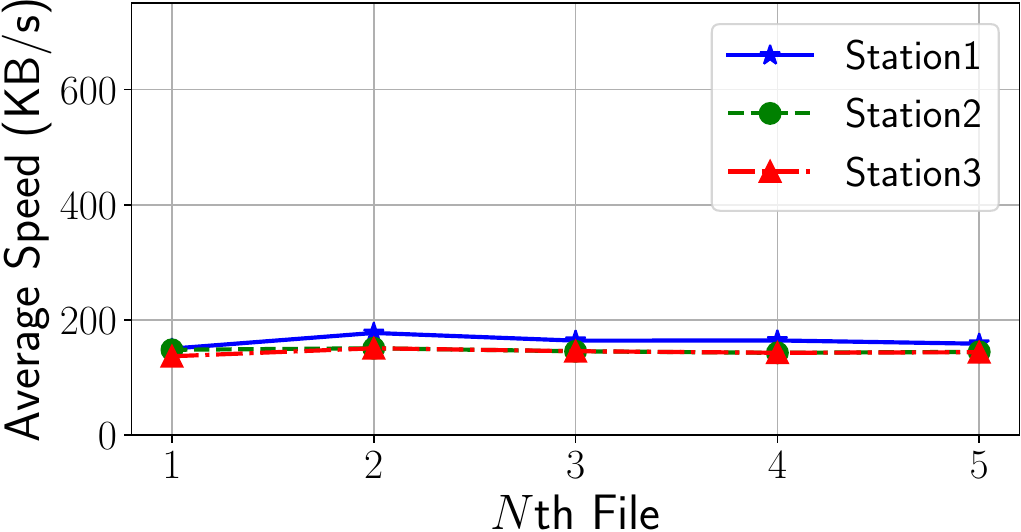}%
		\caption{}%
		\label{copy_to_server_off_vanila}
	\end{subfigure}\hspace{2em}
	\begin{subfigure}[b]{0.45\columnwidth}
		\includegraphics[width=\columnwidth]{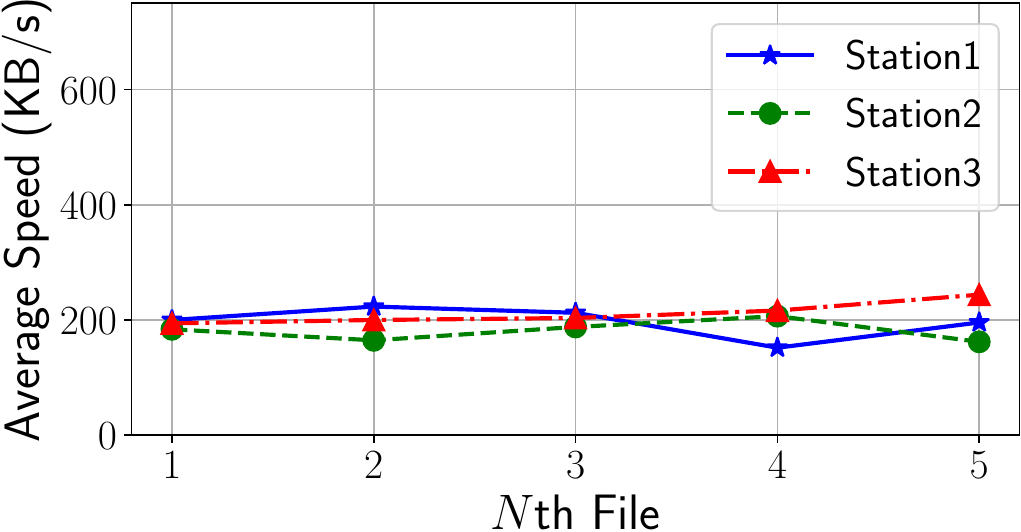}%
		\caption{}%
		\label{copy_to_server_off}
	\end{subfigure}
	
	\begin{subfigure}[b]{0.45\columnwidth}
		\includegraphics[width=\columnwidth]{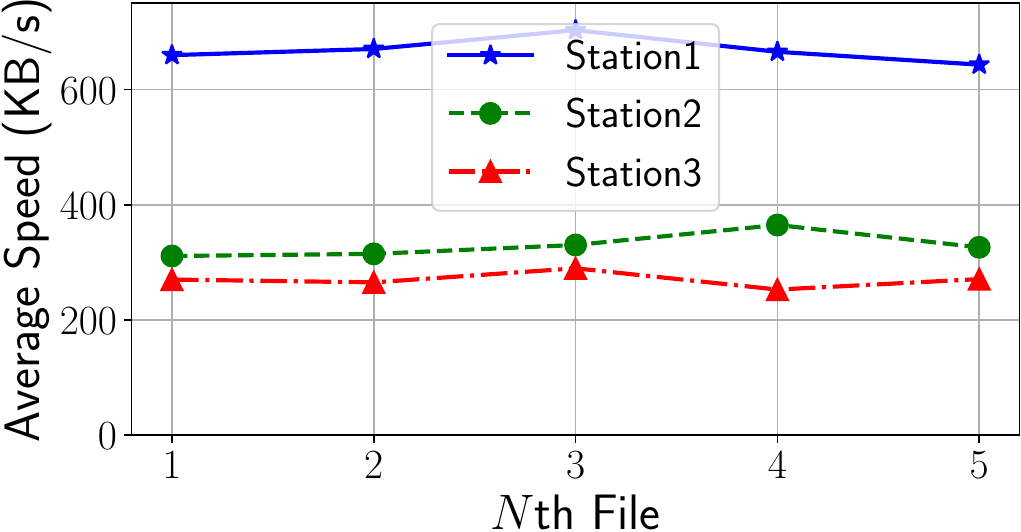}%
		\caption{}%
		\label{copy_to_server_500_vanila}
	\end{subfigure}\hspace{2em}
	\begin{subfigure}[b]{0.45\columnwidth}
		\includegraphics[width=\columnwidth]{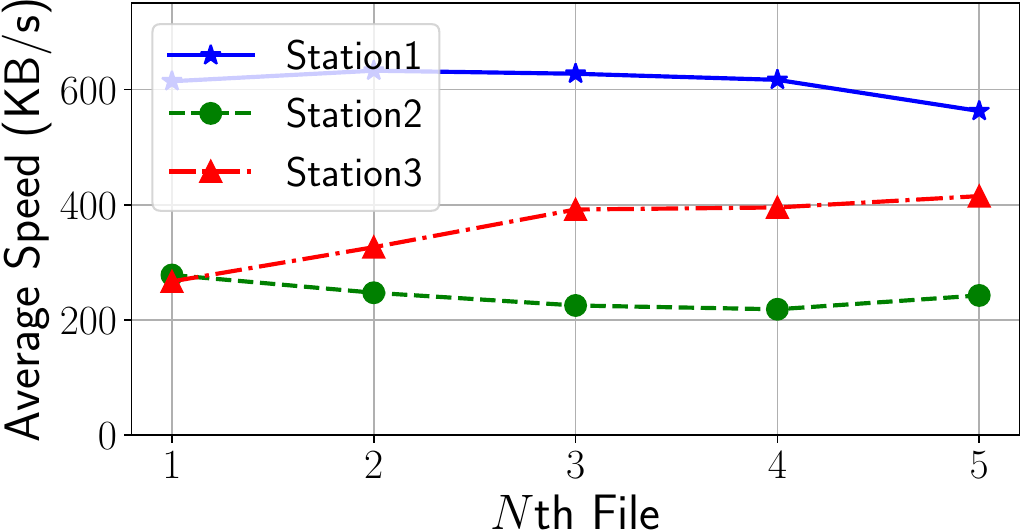}%
		\caption{}%
		\label{copy_to_server_500}
	\end{subfigure}

	\caption{(\subref{copy_to_server_off_vanila}, \subref{copy_to_server_500_vanila}) Vanilla driver and (\subref{copy_to_server_off}, \subref{copy_to_server_500}) our scheme. Average copy speed of 50 $MB$ files to the server (public IP) with (\subref{copy_to_server_off_vanila}, \subref{copy_to_server_off}) RTS/CTS off and (\subref{copy_to_server_500_vanila}, \subref{copy_to_server_500})  RTS\_threshold=500 $bytes$.}
	\label{fig:copy_to_server}
\end{figure}

\Cref{fig:copy_to_server} shows the results for repeatedly copying (uploading) a 50 $MB$ file from the stations to a local server with a public IP address.
We see the results are similar for the two driver versions, while enabling RTS/CTS improves the performance.
The improvement is more evident for $Station_1$, which is in line-of-sight of the AP.

\begin{figure}[t]
	\centering
	\begin{subfigure}[b]{0.45\columnwidth}
		\includegraphics[width=\columnwidth]{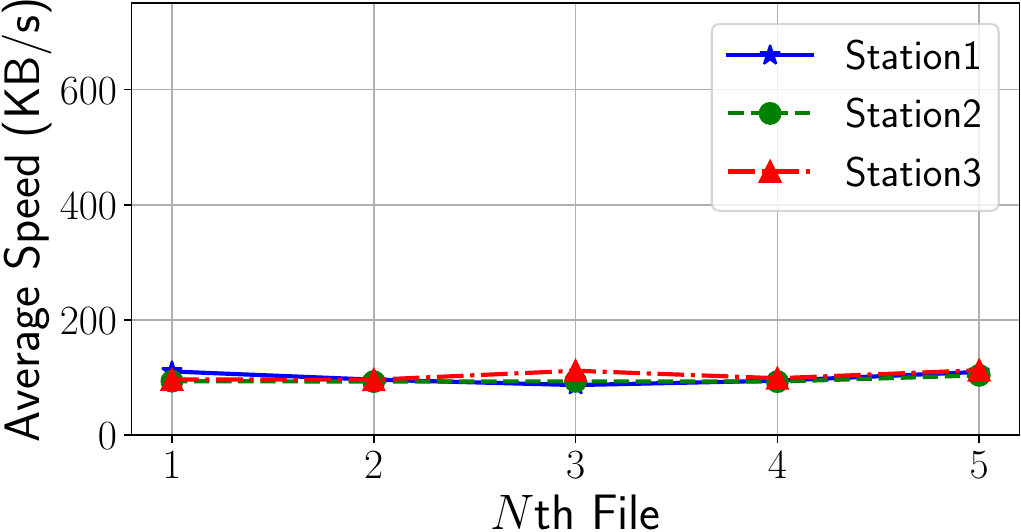}%
		\caption{}%
		\label{copy_from_server_off_vanila}
	\end{subfigure}\hspace{2em}
	\begin{subfigure}[b]{0.45\columnwidth}
		\includegraphics[width=\columnwidth]{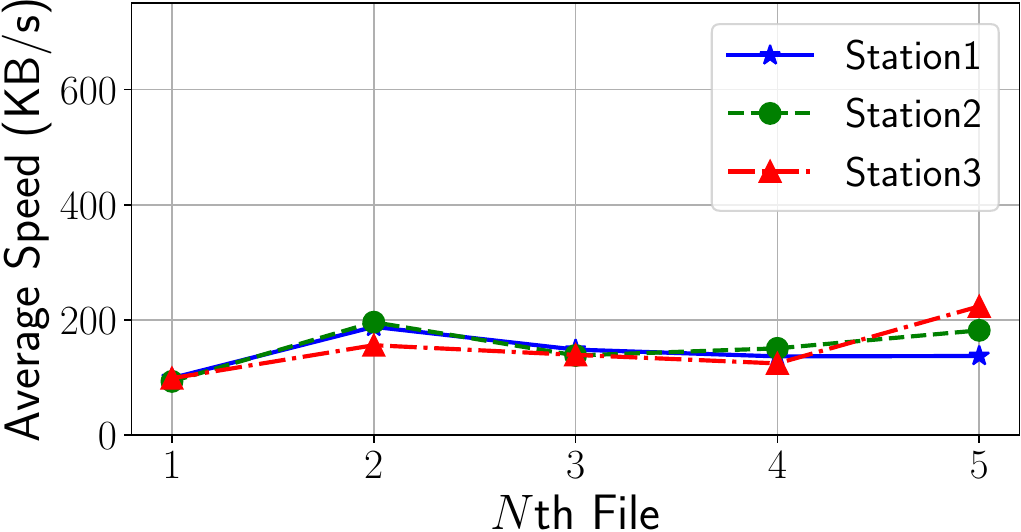}%
		\caption{}%
		\label{copy_from_server_off}
	\end{subfigure}
	\caption{(\subref{copy_from_server_off_vanila}) Vanilla driver and (\subref{copy_from_server_off}) our scheme. Average copy speed of 50 $MB$ files from the server (public IP).}
	\label{fig:copy_from_server}
\end{figure}

\begin{figure}[t]
	\centering
	\begin{subfigure}[b]{0.45\columnwidth}
		\includegraphics[width=\columnwidth]{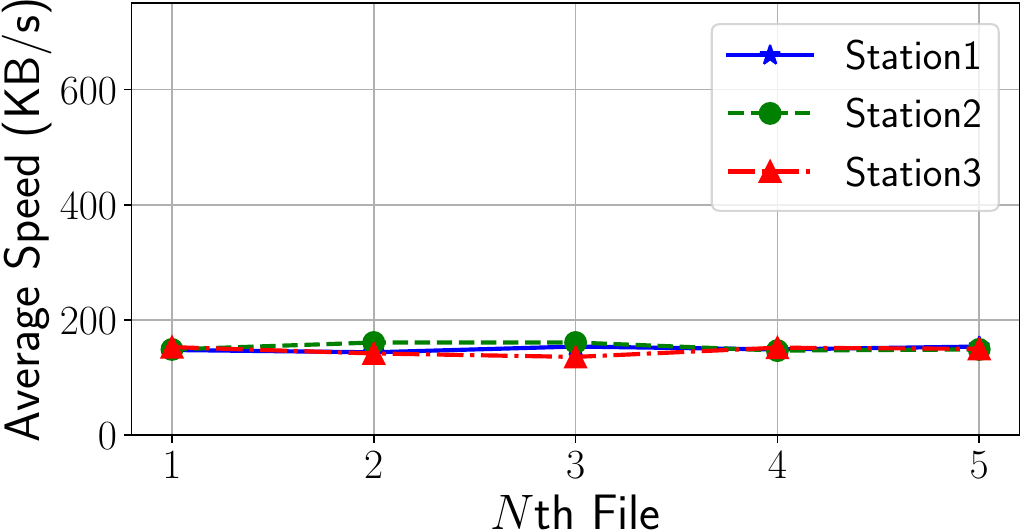} 
		\caption{}%
		\label{wget_off_vanila}
	\end{subfigure}\hspace{2em}
	\begin{subfigure}[b]{0.45\columnwidth}
		\includegraphics[width=\columnwidth]{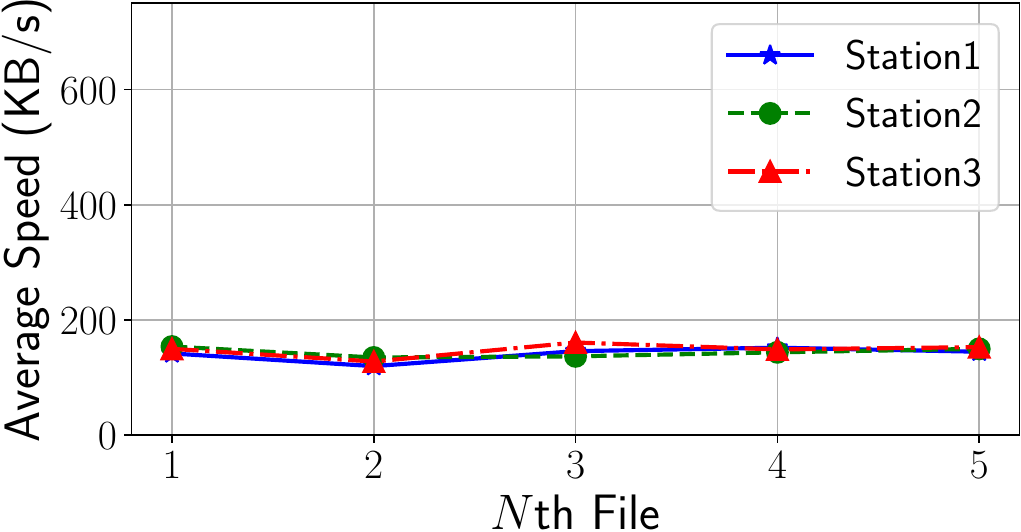}%
		\caption{}%
		\label{wget_off}
	\end{subfigure}
	\caption{(\subref{wget_off_vanila}) Vanilla driver and (\subref{wget_off}) our scheme. Average download speed of 60 $MB$ images from the official Debian website.}
	\label{fig:wget}
\end{figure}

\Cref{fig:copy_from_server} shows the results for stations copying (downloading) a 50 $MB$ file from the same local server.
In this scenario, we didn't enable RTS/CTS on the stations because the majority of heavy transmissions are from the AP to the stations.
The results are again similar for the two driver versions.
For completeness in this context, we also evaluate the speed of downloading a 60 $MB$ ISO file\footnote{\url{https://cdimage.debian.org/cdimage/archive/5.0.10/powerpc/iso-cd/debian-5010-powerpc-businesscard.iso}} from the official Debian website over the Internet.
\Cref{fig:wget} shows the results for the two driver versions are similar too.

The goal is achieved: the modified driver performs very similarly to the vanilla driver.
We do not attempt to explain performance results in further detail, because the intent is not to evaluate the latest Wi-Fi hardware performance or an optimized AP (unlike our device that acted as AP).
Extensive performance evaluation is left for future work, after a successful implementation of our scheme on the latest \acp{NIC} that support IEEE 802.11 n/ac/ax (Wi-Fi 4/5/6).

%% file: section/conclusions.tex
\section{Conclusion and Future work}
\label{sec:conclusion}

Our scheme re-randomizes MAC addresses while the stations are connected to the AP without disrupting ongoing communication.
Synchronized MAC address transitions make linking based on timing information difficult, with the size of the anonymity set being the number of connected stations.
\acp{SN} and nonces are also reset to ensure unlinkability.
We show the feasibility of our scheme with a set of small-scale experiments.
The benefit is a very significant reduction of user exposure to any eavesdropper: any long-lived connection/session with an AP essentially dissolves into a multitude of unlinkable ephemeral connections.

An immediate next step is to implement and evaluate our scheme on the latest \acp{NIC}, with the challenge of using non-free binary firmware.
The $T$ value of our scheme is currently hard-coded into the driver; an extension is to have connected mobile stations receive the re-randomization schedule from the AP.
Our scheme highly depends on time synchronization, especially for high bit rates (i.e., short frame intervals), otherwise station MAC address transitions would be easily linkable.
Countermeasures for loosely synchronized devices would be necessary, along with a full-blown unlinkability analysis.
Adaptation of our scheme in Wi-Fi ad-hoc mode is also a part of future work.